\documentclass[aps,preprint]{revtex4}%
\usepackage{amsfonts}
\usepackage{amsmath}
\usepackage{amssymb}
\usepackage{graphicx}%
\begin{document}
\title{\bigskip{\LARGE ON A TIME VARIATION OF NEUTRINO'S MASS }}
\author{Marcelo Samuel Berman$^{\ast}$}
\affiliation{$^{\ast}$Editora Albert Einstein Ltda.}
\affiliation{Av. Candido Hartmann, 575 - \# 17 - Ed. Renoir }
\affiliation{80730-440 - Curitiba - PR - Brazil}
\author{Fernando de Mello Gomide$^{\ast\ast}$}
\affiliation{$^{\ast\ast}$Departamento de F\'{\i}sica}
\affiliation{Instituto Tecnol\'{o}gico de Aeron\'{a}utica (ITA)}
\affiliation{Centro T\'{e}cnico Aeroespacial }
\affiliation{12200-970- \ S\~{a}o Jos\'{e} dos Campos - SP - Brazil}
\begin{abstract}
After the introduction to the subject, we review the Machian relations
involving the different kinds of energy densities in the Universe. By suposing
that dark-matter is composed by Neutrino's massive energy, \ we hint that the
Neutrino's masses should be proportional to \ $R^{-1}$, while the total number
of Neutrinos in the Universe increases with \ $R^{2}$\ , where \ $R$\ \ stands
for Hubble's length. This makes possible that the dark mater-energy density be
of Neutrinos origin, to vary with \ $R^{-2}$\ , as is necessary for a Machian
Universe. The neutrino average mass is estimated for the present Universe
(=10$^{-76}$g).

\ \ \ \ \ \ \ \ \ \ All correspondence should be sent to the first author.



\end{abstract}
\maketitle

\begin{center}
\bigskip{\LARGE ON A TIME VARIATION OF NEUTRINO'S MASS}
\end{center}

\bigskip

\begin{center}
Marcelo Samuel Berman and Fernando de Mello Gomide
\end{center}

\bigskip

{\Large I - INTRODUCTION}

\bigskip

\bigskip The subject of mass-varying neutrinos, has been very recently given
attention[22][23][24]. But, first, let us make an historical review of the
neutrino research.

In 1930 the Austrian theoretical physicist Wolfgang Pauli, Nobel Prize due to
his research on the electron spin, verified in the processes of electron
emission from neutrons, that the energy-momentum balance was violated. In
other terms: the much solid principle of energy-momentum conservation of
special relativity seemed to loose its preeminence at the basis of every
physical theory. Well something highly perplexing, for in spite of being
universally applied to all known high energy processes, it could not be
corroborated in this special process of electron emission from neutrons, the
so called Beta emission process. A painful problem of course, but brilliantly
solved through Pauli's revolutionary hypothesis and Enrico Fermi's theory, in
1934, [2]. Pauli postulated the existence of a new particle, the neutrino[1]
that was responsible for the disappearence of energy-momentum in the neutron
decay (beta emission). Fermi forwarded the neutrino theory wherein it is
deduced that this new postulated neutral particle has a probability of
interaction with other particles, 10 to power 22 times smaller than what was
known then. Therefore the neutrino acted as if a phantom particle that could
not be easely detected. Only in 1953, twenty years after Fermi's theory, was
the particle to emerge to existence, thanks to the experiments done by two
American experimental physicists, Frederick Reines and Clyde Cowan. They used
then a new detection technique that afforded the possibility for detection of
an effect predicted by Fermi. It was the tecnique based on a different type of
particle detector, the scintillation counter. Being faithful to a fundamental
physical principle, the energy-momentum conservation, Pauli and Fermi were
responsible for one of the most important scientific discovery of the XXth
century. Fermi's predictions from his theory were widely verified by the
community of high-energy physicists, such that nobody doubted the existence of
the putative neutrino. Frederick Reines who received the Nobel Prize in 1995
made the following important epistemological comment:

\textquotedblleft It must be recognized, however, that independent of the
observation of a `free neutrino' interaction with matter, the Fermi theory was
so attractive in its explanation of beta decay that belief in the neutrino as
a `real' entity was general\textquotedblright.[1]

Reines' comment is a remarkable epistemological judgement in so far as to view
certain experimental evidence of a theory, not something clinching or crucial.
Fermi's theory offered a gallaxy of effects in high-energy physics of so
important level, that the putative neutrino could be kept as such for 20 years
without bothering physicists. Reiner and Cowan's discovery of the free
neutrino was not a crucial test of a theory, but an experimental success after
a history of theoretical and other experimental evidences. We are going to
deal afterwards on the question of clinching experimental proofs. An analogous
situation we are going to point out in the history of the heliocentric theory,
accepted by every astronomer of the XVIIth century without the
\textquotedblleft clinching proof\textquotedblright\ of stellar parallax
obtained by Bessel on the XIXth century. Let us remind that Enrico Fermi's
theory of 1934 is considered the birth of quantum field theory[2], which is
cornerstone in elementary particles theory. Fermi received the Nobel in 1938.

The present theory of material reality's offers three families of elementary particles:

a)\qquad6 quarks and 6 antiquarks.

b)\qquad Leptons and antileptons: electron, muon, tau-particle, neutrino-e,
neutrino-mu and neutrino-tau.

c)\qquad Gauge bosons: photon, W+, W-, Z, and Gluon. Being bosons they have no
antiparticles' counterpart.

Let us add some considerations as to the three neutrinos. The neutrino
developed in Fermi's theory and detected by Reines and Cowan was called by
Feinberg in 1958, neutrino-e, since it was related to the beta decay of the
neutron. Besides Feinberg predicted that the neutrino generated by the muon
decay should be different and it was then named neutrino-mu. Through an
experiment done in 1962 by Leon Lederman, Schwartz, and Steinberger, the
prediction was brilliantly corroborated[3]. In 21/7/2000 a group of 54
American, Japanese, Korean, and Greek physicists at Fermilab, announced the
detection of the neutrino-tau.[4] The neutrino-tau results from the decay of
the lepton tau particle. Thus, from 1953 to 2000 the 3 neutrinos predicted in
the theoretical structure had been detected.

From 1967 to 1998 several experiments were done in order to measure the
neutrino flux produced in the Sun. As we know the thermonuclear reactions
inside our star, would produce abundant torrents of neutrinos-e. The first
measurements done in 1967 by Ray Davies offered a perplexing result: the flux
measured had been about 1/3 of the predicted value[5][6]. In 1969 the Italian
physicist Pontecorvo suggested a hypothesis that would account for the missing
Solar neutrinos: en route to Earth part of them would oscillate to a different
type of neutrinos. This hypothesis was endorsed in 1985 by two Russian
physicists Miteyev and Smirnov: from the Sun to Earth a substantial part of
neutrinos-e would transform into neutrinos-mu or tau[7].

According to the standard model of elementary particles theory, the property
of oscillation of a neutrino into another type pressuposes the existence of
mass. Hence the old assumption of the massless neutrino should be disposed
off. A team of 300 American and Japanese physicists at Takayma proceeded to an
experiment with the huge detector Super-Kamiokande in order to detect
oscillation of neutrino-mu into neutrino-tau in the flux of secondary cosmic
rays. In spite of the cautious language of Yoshiro Suzuki of the University of
Tokio, the results strongly favour the oscillation of neutrinos-mu[8]. These
findings with the Super-Kamiokande are regarded as compelling but not
definitive[9]. More recently, Tsuyoshi Nakaya of Kyoto University announced in
June, 2004 at a meeting in Paris, a reassuring confirmation of the oscillation
mechanism, that is: strong evidence of oscillation of neutrino-mu into
neutrino tau[10]. The square masses of the neutrino eigenstates were measured
from the Super-Kamiokande data and were in agreement with the oscillation
parameters from atmospheric data.

Notwithstanding the successes obtained, new theoretical problems are being
tackled in the domain of the physics of neutrinos[11]. Well there is no end in
the quest for knowledge and understanding of reality: problems solved open the
doors for new problems.

\bigskip

{\Large II \ MACHIAN RELATIONS FOR DARK MATTER ENERGY}

\bigskip

\bigskip It has been asserted, that 67\% of the energy density of the
Universe, is due to a cosmological "constant" energy. The restant energy
density is fractioned in two parts: 5\% as visible mass and 28\% as dark
matter. Let us suppose that dark matter is constituted by neutrinos with
non-zero rest mass. Berman[12][13][14] has suggested that, \ if Mach's
principle is understood as meaning that the total energy of the Universe is
null, and if each particular energy contribution to the total energy density,
has constant participation during the whole history of the Universe, one may
obtain different Machian relations. These Machian relations, of which, the
Brans-Dicke [15] relation is a particular case, should not, according to
Berman, be viewed as just coincidental with the present Universe. Suppose that
the total energy is given by:

\bigskip

$E=Mc^{2}-\frac{GM^{2}}{2R}+4\pi\Lambda\frac{R^{3}}{3\kappa}+\frac{L^{2}%
}{MR^{2}}$ \ \ \ \ \ \ \ \ \ \ \ \ \ \ \ \ \ \ \ ,\ \ \ \ \ \ \ \ \ \ \ \ \ \ \ \ \ \ \ \ \ \ \ \ \ \ \ \ \ (1)

\bigskip

where the four terms to the right of relation (1) represent respectively the
inertial, gravitational, cosmological constant's and rotational energies.

\bigskip

When we impose,

\bigskip

$E=0$ \ \ \ \ \ \ \ \ \ \ \ \ \ \ \ \ , \ \ \ \ \ \ \ \ \ \ \ \ \ \ \ \ \ \ \ \ \ \ \ \ \ \ \ \ \ \ \ \ \ \ \ \ \ \ \ \ \ \ \ \ \ \ \ \ \ \ \ \ \ \ \ \ \ \ \ \ \ \ \ \ \ \ \ \ \ \ \ \ (2)

\bigskip

we obtain:

\bigskip

$\frac{GM}{2c^{2}R}-\frac{4\pi}{3\kappa}\left[  \frac{\Lambda R^{3}}{c^{2}%
M}\right]  -\frac{L^{2}}{c^{2}M^{2}R^{2}}=1$ \ \ \ \ \ \ \ \ \ \ \ . \ \ \ \ \ \ \ \ \ \ \ \ \ \ \ \ \ \ \ \ \ \ \ \ \ \ \ \ \ \ \ \ \ \ \ \ \ (3)

\bigskip

If no one of the above terms will increase or decrease different than the
others, we can solve equation (3) by imposing that:

\bigskip

\bigskip$\frac{GM}{2c^{2}R}=\gamma_{1}$ \ \ \ \ \ \ \ \ \ \ \ , \ \ \ \ \ \ \ \ \ \ \ \ \ \ \ \ \ \ \ \ \ \ \ \ \ \ \ \ \ \ \ \ \ \ \ \ \ \ \ \ \ \ \ \ \ \ \ \ \ \ \ \ \ \ \ \ \ \ \ \ \ \ \ \ \ \ \ \ \ (4)

\bigskip$\frac{4\pi}{3\kappa}\left[  \frac{\Lambda R^{3}}{c^{2}M}\right]
=\gamma_{2}$ \ \ \ \ \ \ \ \ \ \ \ , \ \ \ \ \ \ \ \ \ \ \ \ \ \ \ \ \ \ \ \ \ \ \ \ \ \ \ \ \ \ \ \ \ \ \ \ \ \ \ \ \ \ \ \ \ \ \ \ \ \ \ \ \ \ \ \ \ \ \ \ \ (5)

$\frac{L^{2}}{c^{2}M^{2}R^{2}}=\gamma_{3}$ \ \ \ \ \ \ \ \ \ \ \ , \ \ \ \ \ \ \ \ \ \ \ \ \ \ \ \ \ \ \ \ \ \ \ \ \ \ \ \ \ \ \ \ \ \ \ \ \ \ \ \ \ \ \ \ \ \ \ \ \ \ \ \ \ \ \ \ \ \ \ \ \ \ \ \ \ \ \ (6)

\bigskip

where the \ $\gamma$ 's\ \ obey the conditions:

\bigskip

1) \ \ $\gamma_{i}=$ \ constant\ \ \ \ ( \ $i=1,2,3$\ \ )\ \ \ \ \ \ \ , \ \ \ \ \ \ \ \ \ \ \ \ \ \ \ \ \ \ \ \ \ \ \ \ \ \ \ \ \ \ \ \ \ \ \ \ \ \ (7)

\bigskip

2) \ \ \ $\gamma_{1}-\gamma_{2}-\gamma_{3}=1$
\ \ \ \ \ \ \ \ \ \ \ \ \ \ \ \ \ \ \ \ \ \ \ \ \ . \ \ \ \ \ \ \ \ \ \ \ \ \ \ \ \ \ \ \ \ \ \ \ \ \ \ \ \ \ \ \ \ \ \ \ \ \ \ \ (8)

\bigskip

It can be checked that, due to the original Brans-Dicke relation [15],

\bigskip

$\frac{GM}{c^{2}R}\cong1$ \ \ \ \ \ \ \ \ , \ \ \ \ \ \ \ \ \ \ \ \ \ \ \ \ \ \ \ \ \ \ \ \ \ \ \ \ \ \ \ \ \ \ \ \ \ \ \ \ \ \ \ \ \ \ \ \ \ \ \ \ \ \ \ \ \ \ \ \ \ \ \ \ \ \ \ \ \ \ \ \ \ \ \ \ (9)

\bigskip

and also because of the above arguments, all the $\gamma$ 's\ \ have a near
unity value. We thus obtain, with Berman[14], the following variation laws:

\bigskip

$M\propto R$ \ \ \ \ \ \ \ \ ; \ \ \ \ \ \ \ \ \ \ \ \ \ \ \ \ \ \ \ \ \ \ \ \ \ \ \ \ \ \ \ \ \ \ \ \ \ \ \ \ \ \ \ \ \ \ \ \ \ \ \ \ \ \ \ \ \ \ \ \ \ \ \ \ \ \ \ \ \ \ \ \ \ \ \ \ \ (10)

\bigskip

$L\propto R^{2}$ \ \ \ \ \ \ \ \ ; \ \ \ \ \ \ \ \ \ \ \ \ \ \ \ \ \ \ \ \ \ \ \ \ \ \ \ \ \ \ \ \ \ \ \ \ \ \ \ \ \ \ \ \ \ \ \ \ \ \ \ \ \ \ \ \ \ \ \ \ \ \ \ \ \ \ \ \ \ \ \ \ \ \ \ \ \ (11)

\bigskip

$\Lambda\propto R^{-2}$ \ \ \ \ \ \ \ \ . \ \ \ \ \ \ \ \ \ \ \ \ \ \ \ \ \ \ \ \ \ \ \ \ \ \ \ \ \ \ \ \ \ \ \ \ \ \ \ \ \ \ \ \ \ \ \ \ \ \ \ \ \ \ \ \ \ \ \ \ \ \ \ \ \ \ \ \ \ \ \ \ \ \ \ \ (12)

\bigskip

The Machian relations (4)(5)(6) have been noticed long ago, as approximate
relations for the present Universe; the radical departure made by Berman, is
contained in the fact that the \ $\gamma$ 's\ \ are constant during the
lifespan\ \ of the Universe, and not only for the present time, so that
relations (10)(11)(12), are valid during all times.

\bigskip

We now can obtain the corresponding energy densities for the above relations:

\bigskip

$\rho_{1}=\frac{M}{\frac{4}{3}\pi R^{3}}=\left[  \frac{6\gamma_{1}c^{2}}{4\pi
G}\right]  R^{-2}$ \ \ \ \ \ \ \ \ , \ \ \ \ \ \ \ \ \ \ \ \ \ \ \ \ \ \ \ \ \ \ \ \ \ \ \ \ \ \ \ \ \ \ \ \ \ \ \ \ \ \ \ \ \ \ \ \ (13)

\bigskip

$\rho_{2}=\frac{\Lambda}{\kappa}=\left[  \frac{\gamma_{2}\gamma_{1}c^{4}}{2\pi
G}\right]  R^{-2}$ \ \ \ \ \ \ \ \ , \ \ \ \ \ \ \ \ \ \ \ \ \ \ \ \ \ \ \ \ \ \ \ \ \ \ \ \ \ \ \ \ \ \ \ \ \ \ \ \ \ \ \ \ \ \ \ \ \ \ \ (14)

\bigskip

and,

\bigskip

$\rho_{3}=\frac{\left[  \frac{L^{2}}{MR^{2}}\right]  }{\frac{4}{3}\pi R^{3}%
}=\left[  \frac{3\gamma_{1}\gamma_{3}c^{4}}{2\pi G}\right]  R^{-2}$
\ \ \ \ \ \ \ \ . \ \ \ \ \ \ \ \ \ \ \ \ \ \ \ \ \ \ \ \ \ \ \ \ \ \ \ \ \ \ \ \ \ \ \ \ \ \ \ \ \ \ \ \ (15)

\bigskip

We can check that all energy densities are proportional to \ $R^{-2}$\ , so
that, we can also write:

\bigskip

$\rho_{TOT}=$\ \ $\rho_{1}+\rho_{2}+\rho_{3}=\Gamma R^{-2}$ \ \ \ \ \ \ (
\ $\Gamma$\ \ = constant\ \ ) \ \ . \ \ \ \ \ \ \ \ \ \ \ \ \ \ \ \ (16)

\bigskip

In the spirit of inflationary Cosmology [16], we identify, for the present
Universe, \ $\rho_{TOT}$\ \ with the critical density, so that we would have:

\bigskip

$\rho_{TOT}\cong2$\ x $10^{-29}$ g / cm$^{3}$ \ \ \ \ \ \ \ .

\bigskip

In the next Section, we calculate an estimate for neutrinos mass, and its time
variation. But, we observe that, if dark matter is a fraction of \ $\rho
_{TOT}$\ \ , this fraction will also depend on \ $R^{-2}$\ , so as to keep all
relative components\ \ equally balanced\ along time.

\bigskip

{\Large III \ \ \ A theory for neutrinos energy density}

\bigskip

As we have noticed before the energy density of dark matter, to be identified
with neutrinos, shall be given by:

\bigskip

$\rho_{\nu}=0.27\rho_{TOT}$ \ \ \ \ \ . \ \ \ \ \ \ \ \ \ \ \ \ \ \ \ \ \ \ \ \ \ \ \ \ \ \ \ \ \ \ \ \ \ \ \ \ \ \ \ \ \ \ \ \ \ \ \ \ \ \ \ \ \ \ \ \ \ \ \ \ \ \ \ \ \ \ \ \ (17)

\bigskip

Berman [17] along with others (see Sabbata and Sivaran, 1994 [21]) have
estimated that the Universe possess a magnetic field which, for Planck's
Universe, was as huge as \ $10^{55}$\ Gauss. The relic magnetic field of the
present Universe is estimated in \ $10^{-6}$\ Gauss. We can then, suppose that
all neutrinos' spins have been alligned\ \ with the magnetic field. On the
other hand, the spin of the Universe is believed to have increased in
accordance with Machian relation (6) above, which entails relation (11) above.
If we call \ \ $n$\ \ \ the number of neutrinos in the present Universe, and
\ $n_{Pl}$\ \ its value for Planck's Universe, we may write:

\bigskip

$\frac{n}{n_{Pl}}=\frac{L}{L_{Pl}}=10^{120}$\ \ \ \ \ \ \ \ . \ \ \ \ \ \ \ \ \ \ \ \ \ \ \ \ \ \ \ \ \ \ \ \ \ \ \ \ \ \ \ \ \ \ \ \ \ \ \ \ \ \ \ \ \ \ \ \ \ \ \ \ \ \ \ \ \ \ \ \ \ (18)

\bigskip

Then, \ \ \ 

\bigskip

$n=n_{Pl}\left[  \frac{R}{R_{Pl}}\right]  ^{2}$ \ \ \ \ \ \ \ \ \ \ \ \ \ . \ \ \ \ \ \ \ \ \ \ \ \ \ \ \ \ \ \ \ \ \ \ \ \ \ \ \ \ \ \ \ \ \ \ \ \ \ \ \ \ \ \ \ \ \ \ \ \ \ \ \ \ \ \ \ \ \ \ \ \ \ \ (19)

\bigskip

We have just obtained the relation for the increase of the number of neutrinos
with \ $R^{2}$\ \ .

\bigskip

Now, we write the energy density of neutrinos,

\bigskip

$\rho_{\nu}\cong\frac{nm_{\nu}}{\frac{4}{3}\pi R^{3}}$\ \ \ \ \ \ \ \ \ \ \ \ \ \ \ ,\ \ \ \ \ \ \ \ \ \ \ \ \ \ \ \ \ \ \ \ \ \ \ \ \ \ \ \ \ \ \ \ \ \ \ \ \ \ \ \ \ \ \ \ \ \ \ \ \ \ \ \ \ \ \ \ \ \ \ \ \ \ \ \ \ \ \ \ (20)

\bigskip

where \ \ \ $m_{\nu}$\ \ is the rest mass of the average neutrino.

\bigskip

If we impose relation (17)\ \ and simultaneously, relations (16)\ and (20), we
conclude two things:

\bigskip

1$^{st}.$) \ \ $\rho_{\nu}=0.27\rho_{Pl}\left[  \frac{R}{R_{Pl}}\right]
^{-2}$\ \ \ \ \ \ \ . \ \ \ \ \ \ \ \ \ \ \ \ \ \ \ \ \ \ \ \ \ \ \ \ \ \ \ \ \ \ \ \ \ \ \ \ \ \ \ \ \ \ \ \ \ \ \ \ \ (21)

\bigskip

2$^{nd}.$) \ \ $m_{\nu}=\frac{\rho_{Pl}R_{Pl}^{4}}{R}$\ \ \ \ \ \ \ \ . \ \ \ \ \ \ \ \ \ \ \ \ \ \ \ \ \ \ \ \ \ \ \ \ \ \ \ \ \ \ \ \ \ \ \ \ \ \ \ \ \ \ \ \ \ \ \ \ \ \ \ \ \ \ \ \ \ \ \ \ \ \ \ (22)

\bigskip

We see now that while the number of neutrinos in the Universe increases with
\ $R^{2}$\ , the rest mass\ decreases with \ \ $R^{-1}$\ \ ; we may obtain,
with \ \ $R\cong10^{28}$\ cm, that the rest mass of neutrinos should be, in
the present Universe:

\bigskip

$m_{\nu}\cong10^{-76}$\ g\ \ \ \ \ \ \ \ \ \ .\ \ \ \ \ \ \ \ \ \ \ \ \ \ \ \ \ \ \ \ \ \ \ \ \ \ \ \ \ \ \ \ \ \ \ \ \ \ \ \ \ \ \ \ \ \ \ \ \ \ \ \ \ \ \ \ \ \ \ \ \ \ \ \ \ \ \ \ \ \ \ (23)

\bigskip

\bigskip One of us (F.M. Gomide, [25]), has estimated the mass of neutrinos a
long time ago, finding, in a seminal paper , the value, \ $10^{-65}$\ g.
Gomide[25] in fact has equated \ \ $m_{\nu}c^{2}$\ \ with the
self-gravitational energy of the proton. If one equates the self-gravitational
energy of the electron, with \ $m_{\nu}c^{2}$\ , with \ \ $R$\ \ as the
electron Compton radius, we would find \ \ $10^{-72}$\ g. \ \ \ \ \ \ \

\bigskip

\bigskip{\Large IV \ \ Conclusions}

\bigskip

A law of variation for the number of neutrinos in the Universe has been found.
A law of variation for the rest mass of neutrinos was also found.

\bigskip

We remind the reader that Kaluza-Klein's cosmology [18][19], consider time
varying rest masses, in a penta-dimensional space-time-matter, of which the
fifth coordinate is rest mass. The above results can not be rejected, for the
time being, by any known data. \ \ \ We point out, that some of the features
of the present \ calculation, are originated from a seminal paper by Sabbata
and Gasperini [20].

\bigskip

\bigskip{\Large Acknowledgements}

\bigskip

The authors thank M(urari) M(ohan) Som, and are also grateful for the
encouragement by Albert, Paula and Geni, and Luisa Mitiko, Marcelo
Guimar\~{a}es, Nelson Suga, Mauro Tonasse, Antonio Teixeira and Herman J. M. Cuesta.

\bigskip

\bigskip{\Large References}

\bigskip

1.\qquad Franklin, A. (2000) - \textit{The Road to the Neutrino}, Physics
Today, \textbf{53}, N. 2, 22 .

2.\qquad Schweber, S. (2002) - \textit{Enrico Fermi and Quantum
Electrodynamics}, Physics Today, \textbf{55}, N. 6, 31 .

3.\qquad Lederman, L. (1989) - \textit{Observations of Particle Physics from
Two Neutrinos to the Standard Model,} Science, \textbf{224}, 664 .

4.\qquad News Media Contact. Fermilab, 00-12, July 20, 2000.

5.\qquad\textit{Are Neutrinos' Mass Hunters Pursuing a Chimera?}, Science,
\textbf{256}, 731, (1992).

6.\qquad\textit{New Results Yeld no Culprit for Missing Neutrinos}, Science,
\textbf{256}, 1512, (1992).

7.\qquad Schwartzschild, B. (1992) - Physics Today, \textbf{45}, N. 8, 17 .

8.\qquad(idem)(1998) - Physics Today, \textbf{51}, N. 8 .

9.\qquad\textit{Search for Neutrino Mass...} , Science, \textbf{283}, 928, (1999).

10.\qquad\textit{Neutrino Oscillation Has Now Been Seen...} , Physics Today,
\textbf{57}, N. 7, 11, (2004).

11.\qquad Masiero,A.; Vempati,S.K.; Vives,O. (2004) - \textit{Massive
Neutrinos and Flavour Violation}, CERN-PH-TH/2004-142.

12. \ Berman,M.S. (2006) - \textit{Energy of Black Holes and Hawking's
Universe -} in \ \textit{Trends in Black Hole Research,} ed. by Paul Kreitler,
Nova Science, New York.

13. \ Berman,M.S. (2006 a) - \textit{Energy, Brief History of Black Holes, and
Hawking's Universe}, in \textit{New Developments in Black Hole Research,} ed.
by Paul Kreitler, Nova Science, New York.

14. \ \ Berman,M.S. (2006 b) - \textit{On the Machian Properties of the
Universe} - submitted.

15. \ \ Brans, C.; Dicke, R.H. (1961) - Physical Review, \textbf{124}, 925.

16. \ \ Guth, A. (1981) - Phys. Rev. \textbf{D23} , 347.\ \ \ 

17. \ \ Berman,M.S. (2006 c) - \textit{On the Magnetic Field of a Machian
Universe} - submitted.

18. \ \ Wesson, P.S. (1999)\ -\ \ \textit{Space-Time-Matter (Modern Kaluza,
Klein Theory) ,} World Scientific, Singapore.

19. \ \ Berman, M.S.; Som, M.M. (1993) - Astrophysics and Space Science,
\textbf{207}, 105.

20. \ \ Sabbata, V.de; Gasperini, M. (1979) - Lettere al Nuovo Cimento
\textbf{25}, 489.

21. \ \ Sabbata, V.de; Sivaram, C. (1994) - \textit{Spin and Torsion in
Gravitation} , World Scientific, Singapore.

22. \ \ Horvat, R. (2005) - astro-ph/0505507 v2.

23. \ \ Fardon, R.; Nelson A.E.; Weiner, N. (2003) - astro-ph/0309800 v2.

24. \ \ Kaplan, D.B. (2004) - Physical Review Letters \textbf{93}, 091801.

25. \ \ Gomide, F.M. (1963) - Nuovo Cimento \textbf{30}, 672.

\end{document}